\documentclass[superscriptaddress, aps, pra, twocolumn, nofootinbib]{revtex4-2}

\usepackage{dsfont}
\usepackage{upgreek}
\usepackage{amsfonts}
\usepackage{amsmath}
\usepackage{amssymb}
\usepackage{graphicx}
\usepackage{dcolumn}
\usepackage{bm}
\usepackage[colorlinks, allcolors=blue]{hyperref}
\usepackage{braket}
\usepackage{url}
\usepackage{enumitem}
\usepackage[dvipsnames]{xcolor}
\usepackage{soul}
\usepackage{parskip}
\usepackage{apptools}
\usepackage{dsfont}

\definecolor{Red}{rgb}{0.9,0,0}
\definecolor{Blue}{rgb}{0,0,0.9}
\newcommand{\ver}[1]{\textcolor{Red}{#1}} 
\newcommand{\kp}{\ket{\psi}}
\newcommand{\kz}{\ket{0}}
\newcommand{\ku}{\ket{1}}

\newtheorem{definition}{Definition}[section]
\newtheorem{theorem}{Theorem}[section]

\begin{document}

\title{Towards establishing a connection between two-level quantum systems and physical spaces}

\author{V. G. Valle}
\affiliation{Depto. de F\'isica, ICE, Universidade Federal de Juiz de Fora, MG, Brazil}

\author{L. L. Brugger}
\email{ lbrugger@ice.ufjf.br (corresponding author)} 
\affiliation{Depto. de F\'isica, ICE, Universidade Federal de Juiz de Fora, MG, Brazil}

\author{B. F. Rizzuti}
\email{brunorizzuti@ice.ufjf.br}
\affiliation{Depto. de F\'isica, ICE, Universidade Federal de Juiz de Fora, MG, Brazil}

\author{Cristhiano Duarte}
\affiliation{Instituto de Física, Universidade Federal da Bahia, Campus de Ondina, Rua Barão do Geremoabo, s.n.,
Ondina, Salvador, BA 40210-340, Brazil}

\begin{abstract}

\begin{center}
\textbf{Abstract} 
\end{center}

This work seeks to make explicit the operational connection between the preparation of two-level quantum systems with their corresponding description (as states) in a Hilbert space. This may sound outdated, but we show there is more to this connection than common sense may lead us to believe. To bridge these two separated realms---the actual laboratory and the space of states---we rely on a paradigmatic mathematical object: the Hopf fibration. We illustrate how this connection works in practice with a simple optical setup. Remarkably, this optical setup also reflects the necessity of using two charts to cover a sphere. Put another way, our experimental realization reflects the bi-dimensionality of a sphere seen as a smooth manifold. \\     


\end{abstract}

\maketitle

\section{Introduction}
\label{Sec.Intro}


\emph{Quantum phenomena do not occur in a Hilbert space, they occur in the laboratory.} Asher Peres's famous reminder \cite{peres_quantum_2010} clearly stands out among the many ways of dispelling misconceptions about the philosophical content of quantum mechanics. Granted, it is difficult to pinpoint why it is so; is it because of the apparent simplicity and putative form in which it is formulated, or is it because of the hidden profoundness of his interjection?

Putting the explanation of its popularity aside for the moment, the fact we want to emphasise is that Peres's observation accomplishes two things at once. It not only brings to the surface the idea that quantum theory may have a descriptive aspect~\cite{LS13}, but it also dispels the common mistake of equating that descriptive characteristic to the underlying physical reality it is supposed to describe~\cite{Leifer14}. Put another way, quantum theory is so far the best available framework to describe the microscopic world—its predictive power is undeniable~\cite{CB11}. Nonetheless, and here is the caveat, we must stress that it consists of a normative set of assumptions and rules designed to describe and deal with phenomena having no classical explanation.

That is to say that unless we put forward a realist interpretation of quantum mechanics—with its inherent difficulties and that falls short in explaining satisfactorily all the striking quantum phenomena~\cite{APRS13}—the (subjective) take-home message from Peres's remark is that quantum theory is a descriptive theory, that its mathematical framework merely formalises a probabilistic description of certain phenomena that happen in the physical space—very much like in the same spirit of strongly subjectivist interpretations of (classical) probability theory~\cite{Vineberg16,Kemeny55, AA63,Savage54,deF79,vNM44}.


That said, what cannot be forgotten, though, is that there is an intricate connection between the physical space where, in Peres's language, quantum phenomena happen and the Hilbert space where our (probabilistic) description of those phenomena lives in. That is exactly what we set out to do in this contribution. Our introductory contribution intends to fill the gap that is left behind by the ordinary, textbook approaches to quantum physics, which does not bridge the physical space to the Hilbert space—as if the former was just an auxiliary way to motivate and introduce the mathematical machinery of the latter. Using a universal gadget for $SU(2)$ polarization optics \cite{simon_universal_1989}, together with the Hopf fibration \cite{bengtsson_geometry_2017}, we will establish a connection between indistinguishable states in $S^3$ and $S^2 \subset \mathds{R}^{3}$, and interpret this connection beyond the mere mathematical construct.

This work closely follows the scholarship pursued by some of the authors, where we set out to recast, and sometimes reconstruct, mathematical objects that are accepted as `givens' directly from our surrounding physical world \cite{vasconcelos_junior_grandezas_unidimensionais_2018, gaio_grandezas_multidimensionais_2019, rizzuti_operational_topological_2020, rizzuti_is_time_real_line_2022}. In all of those works in this line of inquiry, we have so far kept in mind that our primary target audience is composed of students caught in the crossfire of advanced technical texts and ordinary approaches that, for one reason or another, rush through part of the subtleties involved with the foundations of physics—although we also feel that experienced researchers will also find our works as rewarding auxiliary texts where we introduce novelties while dispelling some missteps.

In this sense, we subdivide this work as follows. In sec.~\ref{Sec.Motivation}, we motivate the connection between the physical space and the Hilbert space through a paradigmatic example we will employ recurrently throughout the text. In sec.~\ref{Sec.RelationPSandSS} we flesh out the mathematical machinery we will use to bridge those two spaces. In sec.~\ref{Sec.UniversalGadget} and sec.~\ref{Sec.PreparingSU2States} we argue that mathematical machinery has a deeper physical meaning. Sec. \ref{Sec.Other} brings out other approaches that have already been considered in the literature concerning geometrical methods for description of quantum systems as well as re-deriving quantum theory from simpler axioms. The last section contains our concluding remarks.

\section{Motivation}
\label{Sec.Motivation}

Our motivation for this approach stems from the fact that two-level pure quantum systems qu\emph{b}its are represented by normalised vectors $\ket{\psi} \in \mathds{C}^2$, and usually parametrised by \cite{nielsen_quantum_2010}
\begin{equation}\label{Eq.ThetaPhiQubit}
    \ket{\psi} = \cos \frac{\theta}{2} \ket{0} + e^{i\varphi} \sin \frac{\theta}{2} \ket{1}.
\end{equation}
Here, $\mathcal{B} = \{\kz, \ku \}$ is the canonical basis of  $\mathds{C}^2$ and $\kp$ lies in the so-called Bloch sphere \cite{simon_hamiltons_1992, arecchi_atomic_1972}.  The parameters $\theta$ and $\varphi$ are borrowed from spherical coordinates, so as usual: $\theta \in [0, \pi]$ and $\varphi \in [0, 2\pi]$. We name it $\mathfrak{S}$ \cite{grossi_one_2023}. This sphere provides a visual, geometrical and intuitive way to analyze qubits: every point on the sphere represents a pure state, the north and south poles are but $\ket{0}$ and $\ket{1}$ and more generally, antipodes represents a pair of orthogonal vectors. One possible way to explore the geometrical connection between $\mathfrak{S}$ and $\mathds{R}^3$ can be established by looking to the density matrix $\rho$ constructed from \eqref{Eq.ThetaPhiQubit}. A direct calculation shows that
\begin{equation}\label{Eq.density.matrix}
    \rho = \kp \bra{\psi} = \frac{1}{2}\left (I + \hat{r}\cdot \Vec{\sigma} \right ),
\end{equation}
where $\Vec{\sigma} = (\sigma_1, \sigma_2, \sigma_3)$ is the collection of the Pauli matrices. As usual, $\sigma_k$, $k=1,2,3$ and we use the representation,
\begin{equation}
    \sigma_1 = \begin{pmatrix}
        0 & 1 \\
        1 & 0
    \end{pmatrix}, \, 
    \sigma_2 = \begin{pmatrix}
        0 & -i \\
        i & 0
    \end{pmatrix}, \, 
\sigma_3 = \begin{pmatrix}
        1 & 0 \\
        0 & -1
    \end{pmatrix}.    
\end{equation}

Moreover, $\hat{r}\in \mathds{R}^3$ in \eqref{Eq.density.matrix}, known as the Bloch vector, so obtained is just the unit vector pointing in an arbitrary radial direction,
\begin{equation}
    \hat{r} = (\sin{\theta}\cos{\varphi},\sin{\theta}\sin{\varphi}, \cos{\theta}).
\end{equation}

Beyond a mere mathematical construction, we will delve into the physical content of the relationship between the physical space—where quantum phenomena are carried out—and the Bloch sphere—where the states involved in those experiments are represented in. 


We emphasise that, for simplicity, throughout the text, we consider only quantum systems described by pure states. In doing so, given any vector
\begin{equation}
    \kp = a \kz + b \ku 
\end{equation}
with complex coefficients $a = x_1 + i x_2$, $b = x_3+ ix_4$, $x_\mu \in \mathds{R}$, $\mu=1,2,3,4$, we demand its normalization, $\braket{\psi \vert \psi} = 1$. Thus, we have
\begin{equation}
    x^2_1+ x^2_2+x^2_3+x^2_4=1.
\end{equation}
That way, $\mathfrak{S}$ is (topologically) identified with the compact sphere $S^3 \subset \mathds{R}^4$ \cite{munkres_topology_1974}. 

Leveraging on an operational procedure, what actually defines a state is its corresponding experimental preparation \cite{jauch_foundations_1973}. Consequently, we anticipate that the parameters $\theta$ and $\varphi$ in equation \eqref{Eq.ThetaPhiQubit} possess both physical and geometrical meaning in terms of a list of procedures in a laboratory. 

Qubits quantum systems are not just toy models or convenient mathematical abstractions belonging exclusively to textbooks. Two-level quantum systems abound in nature, can be easily created in the laboratory \cite{childs_universal_2000, harvey_quantum_2022, unrau_flying_2014}, and are also fundamental building blocks of paradigmatic applications of quantum information and quantum foundations to tasks with no classical analogue \cite{HensenEtAl15,BennettEtAl93,WMPSR10, BB14, Ekert1991} or that provide significant speed-ups and gains in relation to their classical counterparts \cite{AruteEtAl19,AM16,SBBK08, BS17}. However, circling back to Peres's provocation, it is not always the case that a direct a physical interpretation of $\theta$ and $\varphi$ —see eq.\eqref{Eq.ThetaPhiQubit}— in these contexts is readily discerned.

Let us give a concrete example of what we mean by state from our standpoint. In a recent paper \cite{grossi_one_2023}, to give a new purpose to the Stern-Gerlach (SG) experiments, we utilized that traditional setup \cite{stern_weg_1921, gerlach_experimentelle_1921_1, gerlach_experimentelle_1922_2, gerlach_magnetische_1922} as a standard example of how to prepare a qubit as well as to provide a physical meaning to the pair $\theta$ and $\varphi$. These parameters are connected to the orientation of the (non-homogeneous) magnetic field through which the beam of silver atoms passes through. When the beam of atoms is directed into a region permeated by a magnetic field oriented as defined in $\mathds{R}^3$ by $(\theta, \varphi)$, it splits into two branches. Upon selecting one of them, the state of the system is represented by equation \eqref{Eq.ThetaPhiQubit}.

Clearly, the parametrization for pure states on $\mathfrak{S}$ is not unique. In fact, we may consider the entire equivalence class of vectors differing by a constant global phase factor and they are indistinguishable in the sense of providing the same expectation values to any observable. We can rephrase it by saying that pure states can also be represented by orbits of the action of the group $\mathcal{U}(1)$ on $\mathfrak{S}$ defined by standard 
multiplication of vectors by a global phase factor (for more details on group actions and related concepts, see Appendix A). With more details, the orbits are equivalence classes of vectors in $\mathfrak{S}$ differing by a phase, say, $e^{i\alpha} \in \mathcal{U}(1)$. The geometry of indistinguishable states is illuminated by the Hopf fibration \cite{bengtsson_geometry_2017}.  Far from being only a mathematical construct, this fibration explores the intrinsic connection between the dimension of the Hilbert space $\mathds{C}^2$ where qubits's representation live and the dimension of the physical space $\mathds{R}^3$ where the system is prepared—in ref.\cite{grossi_one_2023}, the authors partially address this relationship.

Strengthening this relationship, our primary objective is to uncover the \emph{operational} meaning of the parameters $(\theta, \varphi)$ for concrete examples. This requires finding an experimental prescription that associates these parameters with physical significance, specifically for polarized photons.

To achieve this, we need to take several steps. To begin with, we delve into the details of the Hopf fibration—this more mathematical argument will clear the path for what is to come. Next, we will present a kind of universal device for polarization optics, similarly to what has been proposed in \cite{simon_universal_1989}. This device comprises two half-wave and two quarter-wave plates coaxially mounted, allowing for the generation of any $SU(2)$ polarization transformation. Following that,  we directly apply this device to obtain the parametrization \eqref{Eq.ThetaPhiQubit}, providing a straightforward interpretation of $\theta$ and $\varphi$ in terms of adjustable parameters within the plates.   

\section{The relation between physical space and space of states: Hopf fibration}\label{Sec.RelationPSandSS}

Our starting point comes from the geometry of indistinguishable states on $\mathfrak{S}$. Two state vectors $\kp$ and $\ket{\psi'}$ that exclusively differ by a global phase factor 
\begin{equation}
    \ket{\psi'} = e^{i \alpha} \kp, \, \alpha \in \mathds{R}
\end{equation}
are said to be indistinguishable. Such indistinguishability results from the fact that both states provide the same expectation values of an observable:
\begin{equation}
\bra{\psi'} \mathcal{A}\ket{\psi'} = \bra{\psi} \mathcal{A}\kp,
\end{equation}
where $\mathcal{A}$ denotes a self-adjoint operator representing a given physical observable. In this way, (pure) states can be seen as the equivalence classes, or orbits \cite{nakahara_geometry_1990}, of the action of the group $\mathcal{U}(1)$ on  $\mathfrak{S}$,\footnote{The details of some set theory concepts and action of groups on the latter are presented in the Appendix A.} defined by
\begin{equation}
\kp \mapsto e^{i\alpha} \kp.
\end{equation}
Given a representative $\kp = a \kz + b \ku$, we label its unique  corresponding orbit $[(a,b)] \subset \mathds{C}^2 / \{ (0,0) \}$. Each orbit, in turn, can be identified by the complex number given by the ratio
\begin{align}\label{5.2}
    h: \mathds{C}^2/ \{ (0,0) \} &\rightarrow \mathds{C} \nonumber \\
    [(a,b)] & \mapsto h \left( [(a,b)] \right) = \frac{b}{a}, \, a \neq 0. 
\end{align}
Clearly, $h$ is well-defined for equivalence classes, as its action is independent of the class representative - see Appendix A. 

Since the map $h$ takes its values on the space $\mathds{C}$, which, in turn, can be bijectively mapped onto the sphere $S^2 \subset \mathds{R}^3$, we have a clue for connecting $S^2$ to $S^3 \subset \mathds{R}^4$. First, let us consider the equatorial (bijective) stereographic projection 
\begin{align}
    T: S^2 \subset \mathds{R}^3 &\longrightarrow \mathds{C} \nonumber \\
    (x_1,x_2,x_3) &\longmapsto T(x_1,x_2,x_3) = X + i Y = R e^{i \theta}.
\end{align}
$(x_1, x_2, x_3)$ are the coordinates of a point on the sphere $S^2$, while $X$  and $Y$  are the real and imaginary (resp.) parts of a complex number with magnitude $R$ and phase $\theta$. The inverse map is given by
\begin{align}
    T^{-1}(z = X+iY) = \quad \qquad \qquad \qquad \qquad \qquad \qquad \nonumber \\ = \left ( \frac{2X}{X^2 + Y^2 +1}, \frac{2Y}{X^2 + Y^2 +1}, \frac{X^2+Y^2-1}{X^2+Y^2+1} \right ).
\end{align}

The anticipated connection between the space of states and the physical space is established through the following composition $T^{-1}\circ h: \mathds{C}^2/ \{ (0,0) \} \subset S^3 \rightarrow S^2$, explicitly given by
\begin{equation}\label{proj.fibration}  
    (T^{-1}\circ h)\left ([(a,b)] \right ) = \left ( 2 \mbox{Re}(b a^*), 2 \mbox{Im}(b a^*), \vert b \vert^2 - \vert a \vert^2 \right ).  
\end{equation} 
We emphasize that eq. \eqref{proj.fibration} is a consistent operation for equivalence classes, as the three entries on its r.h.s. depend on combinations of the form $z_1 z_2^*$, which renders it independent of the phase factor for different elements within the class. The details of this specific calculation may be found in \cite{grossi_one_2023}.

\begin{figure}
    \centering
    \includegraphics[scale=0.19]{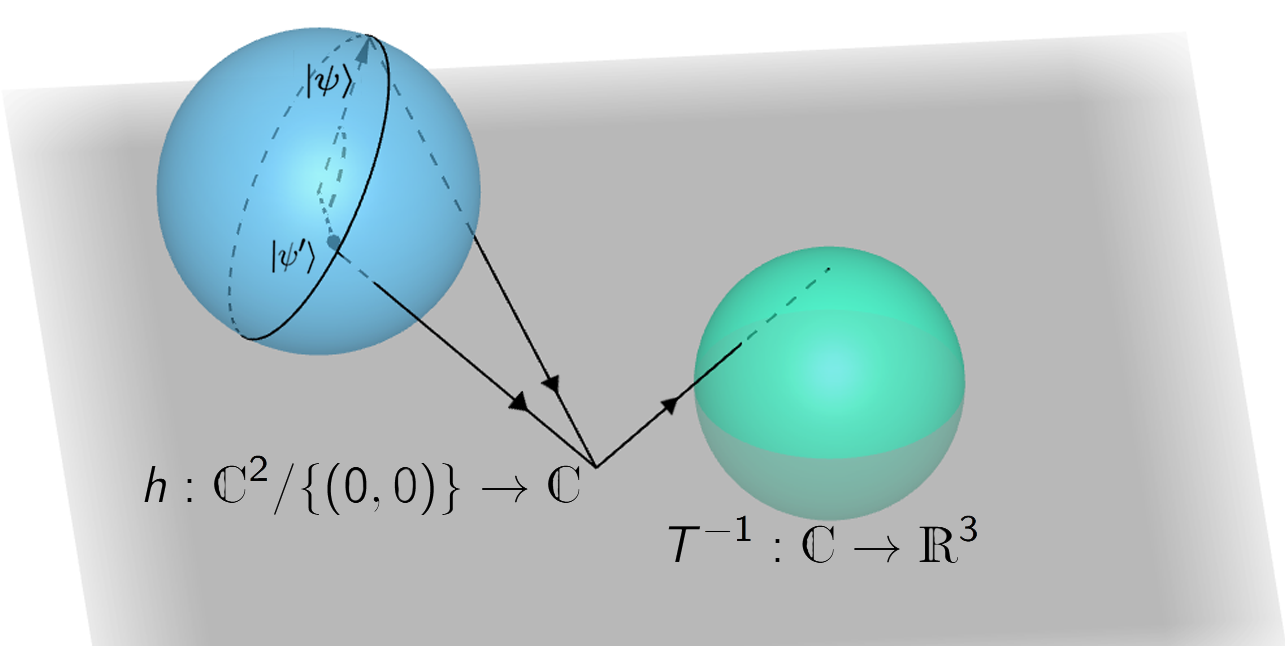}
    \caption{Geometrical representation of the Hopf fibration.}
    \label{fig:hopf.fibration}
\end{figure}

The triple $(T^{-1}\circ h, S^3, S^2)$ defines the Hopf fibration \cite{andre.rafa.impa.2021}. We interpret the composition $T^{-1}\circ h$ as a projection $\pi$ from $S^3$ onto the base space $S^2$. On one hand, indistinguishable states are projected onto the same point $P$ in $S^2$. On the other hand, the inverse image $\pi^{-1}(P)$ contains subsets that correspond to the orbits resulting from the action of $\mathcal{U}(1)$ on $S^3$, which are great circles: $\pi^{-1}(P) \cong S^1$—for the action of a group over a set, see Appendix \ref{appendixA}. In geometric terms, we summarize this as $S^1 \hookrightarrow S^3 \to S^2$. The representation of such construction is depicted in the Fig. \ref{fig:hopf.fibration}.

This mathematical characterization implies that qubit states within $\mathfrak{S} \cong S^3 \subset \mathds{C}^2$ can be prepared in the physical space represented by $S^2 \subset \mathds{R}^3$. 

In ref.\cite{grossi_one_2023}, focusing exclusively on the Stern-Gerlach experimental setup, the authors have established a similar connection between the state space and the physical space. There they give a prescription on how   an ideal Stern-Gerlach apparatus, with a magnetic field pointing in a specific direction
\begin{equation}\label{bloch.vector}
    \hat{r} = (\sin\theta \cos \varphi, \sin \theta \sin \varphi, \cos \theta)
\end{equation}
can be controlled to prepare the generic qubit state given by eq. \eqref{Eq.ThetaPhiQubit}. 

Also in ref. \cite{grossi_one_2023}, the authors have shown how to turn dimension witnesses on their heads so they could be used to bound the dimension of the physical space-instead of bounding a target quantum system’s dimension. The core of their analysis resided on the particularities of the Stern-Gerlach experiment. In this sense, their approach may leave room for criticism for being apparently too system-dependent—more on this issue later. In the present work, although dimensionality does play a fundamental role, we will not focus on dimension witnesses, in other words, we will put aside any probabilistic data to restore the dimensionality of Hilbert spaces. In this sense, the central question we will address here then, is: what implications can we draw for other quantum systems? Put another way, does the Hopf fibration suggest a `universal' approach to preparing states in $\mathds{C}^2$ within a laboratory embedded in physical space represented by $\mathds{R}^3$?

Consider another simple two-level quantum system involving a beam of light passing through a tourmaline crystal, borrowed from Dirac's seminal work \cite{dirac_principles_2010}. In this experiment, it is observed that when the beam is polarized at an angle $\alpha$ to the optic axis of the crystal, only a fraction, $\sin^2 \alpha$, passes through according to classical electrodynamics. In terms of a single photon, this implies a probability of $\sin^2 \alpha$ for it to be found on the back of the crystal, while $\cos^2 \alpha$ represents the probability of absorption.

However oversimplified this description is, the experimental scenario we borrowed from Dirac's study suggests that photons can be described by a state vector in $\mathds{C}^2$, which we will use in a while. This is nothing new, but we needed a concrete and easy-to-digest case to work with. That is exactly what we will do — and we will see that in working with a concrete case, we can devise a truly universal and system-independent argument.

The polarization of an electromagnetic wave can be described in the laboratory through the vector structure of the corresponding electric field. Consider the simplest case of a monochromatic wave with frequency $\omega$ traveling in the $z$ direction. The complete description of polarization can be provided, among others, by the electric field given by
\begin{equation}
    \Vec{E}(t) =\left (E_{0x}\cos(\omega t - \delta_x), E_{0y}\cos(\omega t - \delta_y) \right ).
\end{equation}

In this case, we observe the most general form of elliptical polarization. However, we should emphasize that there is a missing dimension. In the following sense: if we naively equal degrees of freedom to the dimension of environment space, then the preparation of the polarized beam of light is confined to a plane, a two-dimensional space where the electric field resides, despite the fact that it can be immersed in three dimensions, we only need two degrees of freedom to completely describe the polarized light. Yet, in line with the prescription of the Hopf fibration, one would anticipate a three-dimensional background. How can we reconcile these two perspectives? We will address this issue in the forthcoming sections.

\section{Building a universal gadget for polarization of states}\label{Sec.UniversalGadget}

Expanding our previous discussion, we start with the assumption that a beam of polarized photons can be represented by a state in $\mathfrak{S} \subset \mathds{C}^2$. As we seek to describe a preparation of states that not only align with the parametrization in eq.~\eqref{Eq.ThetaPhiQubit}, but that also addresses the missing (background) dimension in the space of state preparation hinted out at the end of the previous section, we turn to a tool known as the `universal gadget for polarization optics' \cite{simon_universal_1989}.

\begin{figure}
    \centering
    \includegraphics[scale=0.4]{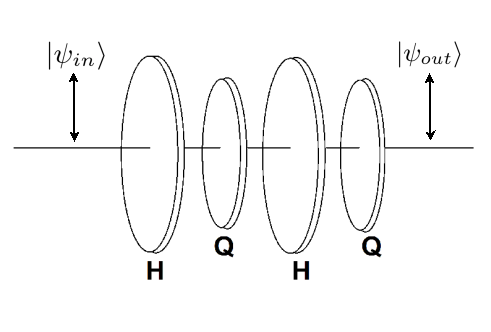}
    \caption{Schematic drawing for the preparation of a state $\ket{\psi_{out}}$ from an arbitrary initial state $\ket{\psi_{in}}$. Note that $\ket{\psi_{in}}$ and $\ket{\psi_{out}}$ represent physical entities actually prepared in the laboratory.}
    \label{fig:inout}
\end{figure}

The universal gadget consists of two half-wave and two quarter-wave plates that are coaxially aligned---see fig.~\ref{fig:inout}. This setup is capable of realizing every $SU(2)$ polarization transformation. This particular group represents linear transformations on field vector components of light beams, keeping the intensity untouched. Thus, $SU(2)$ is key in polarization optics. Leveraging on operational grounds, how can we realize such transformations in a laboratory? Putting aside the technical and experimental intricacies inherent to the realization of this setup in the actual laboratories, we turn our attention to the mathematical structure of this tool. One may use quarter-wave and half-wave plates, which introduce (mathematically) rotations of $\pi/2$ and $\pi$, respectively, as particular instances of $SU(2)$ elements. The gadget generalizes these transformations for any $SU(2)$ polarization. For that, one rotates the plates around the common axis, obtaining, any particular $SU(2)$ polarized state. The corresponding angular positions of the plates not only realize uniquely the group polarization transformation as well as prepare arbitrary states, represented by a vector on the Bloch sphere. 

Both half ($H$) and quarter-wave ($Q$) plates are represented by the conjugation

\begin{align}
    H_\varphi &= \vartheta(\varphi) i \sigma_3 \vartheta (\varphi)^{-1} \\
    Q_\varphi &= \vartheta(\varphi) \begin{pmatrix}
        e^{i \pi/4} & 0 \\
        0 & e^{-i \pi/4}
    \end{pmatrix}\vartheta (\varphi)^{-1}
\end{align}
where $\vartheta (\varphi)$ is the $SO(2)$ element
\begin{equation}
    \vartheta (\varphi) = e^{-i \varphi \sigma_2} = \begin{pmatrix}
        \cos \varphi & - \sin \varphi \\
        \sin \varphi & \cos \varphi 
    \end{pmatrix}.
\end{equation}

The main result from ref.~\cite{simon_universal_1989}, that we are using here, is that any element $U(\xi, \eta, \zeta)$ representing an arbitrary $SU(2)$ polarization transformation can be written as the composition
\begin{align}\label{Eq.QHComposition}
    U(\xi, \eta, \zeta) = Q_{\frac{\xi}{2} + \frac{\pi}{4}}H_{\frac{\xi}{2}+ \frac{\eta}{4} \mp \frac{\pi}{4}}Q_{\frac{\xi}{2} - \frac{\pi}{4}}H_{\frac{\xi - \zeta}{4} \pm \frac{\pi}{4}}.
\end{align}

The parameters $\xi$, $\eta$, $\zeta$ are the known Euler angles, and one can re-write\footnote{Different decompositions of unitary operations are allowed, see for example refs. \cite{nielsen_quantum_2010, tung_group_1985}.} eq.\eqref{Eq.QHComposition} in a neater format:
\begin{equation}
    U(\xi, \eta, \zeta) = e^{-\frac{i}{2}\xi \sigma_2}e^{\frac{i}{2}\eta \sigma_3}e^{-\frac{i}{2}\zeta \sigma_2}.
\end{equation}
According to ref.~\cite{simon_universal_1989}, while $\xi$ and $\zeta$ range in $[0,2\pi]$, $\eta$ is restricted to $[0,\pi]$. As we will argue in the next section, this latter restriction will strongly impact the functioning of our universal gadget, we will need to input at least two states through it to cover the entirety of the Bloch sphere—additionally, the same restriction on $\eta$ will also unexpectedly reflect the fact that spheres are two-dimensional objects. 

These parameters are viewed as adjustable values that an experimentalist can handle within the plates to prepare arbitrary $SU(2)$ polarized states.

The example above is but a concrete proxy of a quantum circuit. Within this context, the word \textit{universal} means that any quantum computation on qubits (be it photons, ion trap, superconductors, quantum dots, etc) can be generated by a finite set of unitary gates. Mathematically, it means to decompose a $SU(2)$ operator in terms of rotations
\begin{equation}
    \begin{pmatrix}
      \cos \alpha/2 & -\sin \alpha/2 \\
\sin \alpha/2 & \cos \alpha/2  
    \end{pmatrix}
\end{equation}
and $\hat{z}$ rotations
\begin{equation}
    \begin{pmatrix}
      e^{-i\beta/2} &  0 \\
0 & e^{-i\beta/2}     
    \end{pmatrix}
\end{equation}
together with a (global) phase shift - see eq. (1.17) of \cite{nielsen_quantum_2010}. Hence, with this prescription, we can obtain an arbitrary quantum logic gate acting on single qubits. What our work provides is an operational meaning to the parameters involved in the qubit state in terms of laboratory tasks for the particular example above of polarization optics.

In the next section, we will show how we can prepare the whole Bloch sphere as an outcome of the universal gadget when we input through it a particular set of states. 

\section{Preparing arbitrary SU(2) state polarization}\label{Sec.PreparingSU2States}

Interpreting the Euler angles as accessible parameters for state preparation enables us to apply $U(\xi, \eta, \zeta)$ to a specific initial state $\ket{\psi_{in}}$ and examine the resulting outcome of this preparation. We express this as follows,
\begin{equation}
    \ket{\psi_{out}} = U(\xi, \eta, \zeta)\ket{\psi_{in}}.
\end{equation}
This can be visualized as allowing the incident state to pass through the universal gadget, as illustrated in Figure \ref{fig:inout}.


As our starting point, we select $\ket{\psi_{in}}=\kz$. We also set~\footnote{The motivation for such choice stems from the fact that any direction in the three-dimensional space $\mathds{R}^3$ can be defined uniquely by the two angles $\theta$ and $\varphi$. Thus, by fixing $\xi=0$, we are left with two parameters, $\eta$ and $\zeta$, to be identified in one way or the other with $\theta$ and $\varphi$.} $\xi=0$. That way, we calculate $\ket{\psi_{out}}$ in the canonical representation of the basis $\{\kz, \ku \}$.
\begin{align}
    \ket{\psi_{out}} &= U(0, \eta, \zeta)\kz = e^{\frac{i}{2}\eta \sigma_3}e^{-\frac{i}{2}\zeta \sigma_2} \kz \nonumber \\
    &= \begin{pmatrix}
        e^{i\frac{\eta}{2}} & 0 \\ 
        0 & e^{-i\frac{\eta}{2}}
    \end{pmatrix}
    \begin{pmatrix}
        \cos \zeta/2 & - \sin\zeta/2  \\
        \sin \zeta/2 & \cos \zeta/2
    \end{pmatrix}
    \begin{pmatrix}
        1 \\ 0
    \end{pmatrix} \nonumber \\ 
    &=\begin{pmatrix}
        e^{i\frac{\eta}{2}}\cos \zeta/2 \\
        e^{-i\frac{\eta}{2}}\sin \zeta/2
    \end{pmatrix}.
\end{align}
Because global phase factors are statistically irrelevant, we can factor out $e^{i\frac{\eta}{2}}$, resulting in the equivalent state
\begin{equation}\label{Eq.ZetaEtaWestern}
    \ket{\psi'_{out}} = \cos \frac{\zeta}{2} \kz + e^{-i\eta}\sin \frac{\zeta}{2} \ku.
\end{equation}

A comparison between eq.~\eqref{Eq.ThetaPhiQubit} and eq.~\eqref{Eq.ZetaEtaWestern} indicates that our universal gadget can, indeed, generate any state on the Bloch sphere's Western hemisphere. We are half-way through the claimed universality of such gadget, as we still need to cover its Eastern hemisphere.

To do so, we similarly select a second initial state $\ket{\psi_{in}}=\ket{1}$ and set $\xi = 0$. With an analogous calculation, we obtain
\begin{equation}\label{Eq.ZetaEtaEasstern}
    \ket{\psi''_{out}} = -\sin \frac{\zeta}{2} \kz + e^{-i\eta}\cos \frac{\zeta}{2} \ku.
\end{equation}

We notice that $\braket{\psi''_{out}\vert \psi'_{out}}=0$, which allows us to conclude that the $\ket{\psi'_{out}}$ and $\ket{\psi''_{out}}$ are antipodes on $\mathfrak{S}$. Since the former lies on the Western hemisphere, the latter is in fact in the Eastern hemisphere, sweeping all the points on the Bloch sphere.  

\section{Other approaches}\label{Sec.Other}

At this stage, what we have done so far should be clear. Assuming quantum theory's textbook abstract formalism and taking Peres' dictum seriously—the fact that quantum phenomena happen in the laboratory, in the physical space we live in—we wanted to investigate and give a physical interpretation to the connection between these two separate worlds. Perhaps what is less clear, and that we could not emphasise enough, is that we are not the first to take this path. This section contains a non-comprehensible review of other works that are similar to the approach we are using here. 
    
Others works seek to explicit the connection between the preparation of qubits an their description in the Hilbert space, focusing on purely geometric grounds, putting aside any operational appeal. In this fashion, the articles \cite{mosseri_geometry_2001, mosseri.qubits.2003} show a Hopf fibration for two an three qubits (respectively) laying on the $S^7$ and $S^{15}$ spheres. As a matter of fact, in both papers, the authors replaced the complex numbers in the one qubit Hopf fibration with quaternions and octonions, essentially expressing two new mappings labeled as $S^7$ and $S^{15}$ fibrations. These apparently simple changes not only had the consequence of allowing a geometric interpretation for two and three qubits, but also showed a interesting fact about the sensitivity of both new mappings when it comes to entanglement of said qubits. So instead of the usual Hopf fibration that was extensively explored on this paper with $S^3$ as the Hilbert space, $S^2$ base and $S^1$ as fibers, we now have, with quaternions, an $S^7$ Hilbert space with $S^4$ base and $S^3$ as fibers, and with octonions, an $S^{15}$ Hilbert space with $S^8$ base and $S^7$ as fibers.

Recall that we are not only building upon Peres' famous reminder but we are also drawing from former instrumental reframings of physical concepts \cite{grossi_one_2023,rizzuti_operational_topological_2020,rizzuti_is_time_real_line_2022}. In this sense, there are  four paradigmatic works that also bear some resemblance to our approach. In ref.~\cite{hardy01}, the author is interested in reconstructing textbook quantum mechanics from physically oriented axioms. Quantum theory is considered a generalization of classical probability theory, and its reconstruction is based on the operational prepare (transform) and measure scenario. With the same perspective, in \cite{dakic_quantum_2009}, quantum theory is reconstructed from three axioms. The former is separated from other probability theories by exhibiting entanglement without contradicting the other axioms. The similarity with the work we do here stops at the operational flavour refs. \cite{hardy01, dakic_quantum_2009} are based on. We are not interested in re-deriving the abstract Hilbert space underlying quantum phenomena, as we remarked earlier, we assume all the usual machinery of quantum theory and want to give an operational meaning to the connection between the abstract and the real world. Similarly, in ref.~\cite{MM11} the authors also re-derive parts of quantum theory from simpler axioms. Their axioms are requirements (informationally motivated) imposed on general probabilistic theories, that in turn are best understood when modelling prepare (transform) and measure scenarios. The same line of thought is also followed by \cite{chiribella_informational_2011}, where a broad class of information-processing operational theories can be traced out in an axiomatic perspective; quantum mechanics is singled out among the class representatives by imposing purification as a postulate. Granted, in all of those works, besides reconstructing parts of quantum theory, they were also able to talk about the dimensionality of the state space for qubits—a topic we marginally touch upon in this work—but we emphasise that this is not what we set out to do here. By giving a physical meaning to the connection between the state space and the physical space, we wanted to start making actual sense of the extra Hilbert space's dimensions squeezed into the three-dimensional physical space we live in—in a rigorous way. Although the low dimensional instance considered here, all other higher-dimensional systems can be reconstructed out of the two-dimensional ones (according to \cite{dakic_quantum_2009}), the latter being considered underlying constituents of the world, reinforcing our focus on low-dimensional systems.

It is undeniable that some aspects of quantum theory are still up for debate, including its own axiomatic formulation and reconstruction through simpler and physically more appealing postulates—besides, obviously, other fundamental and possibly inherent characteristics: Bell non-locality, contextuality, quantum steering, and entanglement to mention a few. What is usually set aside or partially forgotten in many debates about quantum theory is that quantum effects happen in the laboratory, in the physical space we live in, and accordingly each and every piece of mathematical abstraction should find its counterpart in that physical realm. It is exactly this task we are addressing with this contribution—it is exactly this task that sets us apart from the previous works we elaborate on above.     

\section{Conclusion}\label{Sec.Conclusion}

In this paper, we have investigated the intricate relationship between the state space of two-level quantum systems and the physical space in which these systems are prepared. The necessity of such presentation arises when we adopt a universal character of the Hopf fibration in the following sense. By one side the mathematical construct indicates that any two-level quantum system could be prepared in a three-dimensional physical space. A concrete example of this  connection was explored in \cite{grossi_one_2023}, where the authors showed, through a modern rereading of the Stern-Gerlach experiment, how the orientation of the magnetic field in an arbitrary direction in space prepares a qubit state represented by \eqref{Eq.ThetaPhiQubit}. On the other hand, it seems that there is a missing dimension when we think of polarized light as another example of qubit being described by the two-dimensional - and not three - plane where its electric field oscillates. With more details:

1. Using the Hopf fibration, we establish a connection between indistinguishable pure states in $S^3$ and $S^2 \subset \mathds{R}^3$. Our interpretation demonstrates that this connection is more than a mere mathematical construct. Rather, we view the fibration as a description of how states are prepared, with adjustable parameters in a laboratory defining the Bloch vector \eqref{bloch.vector}. The corresponding inverse image leads to the states parameterized by \eqref{Eq.ThetaPhiQubit}.

2. While the fibration does not discriminate among specific two-level quantum systems, we have chosen to focus on an illustrative example in our work: polarized light. Initially, one might think that the plane of oscillation of the electric field fully specifies polarization. However, this notion contradicts the structure revealed by the fibration, which implies a three-dimensional space of preparation. To reconcile this contradiction, we employed a universal gadget for $SU(2)$ polarization optics. Our investigation demonstrated that the adjustable parameters within this device define a state in the Bloch sphere. Consequently, we have successfully ascribed a practical, operational interpretation to the parametrization \eqref{Eq.ThetaPhiQubit}, aligning with our objectives.

To conclude, we stress that in our manuscript we were more concerned with the operationalization and preparation of states in the laboratory, instead of focusing in technical details such as interaction of spin and external magnetic field in the case of Stern-Gerlach setup or even interaction of electric field of the photon and the electric dipole moment implied by the charges of a polarizing plate. The consistency of maintaining the ``prepare and certify'' structure, present in both the sequential combination of two SG's, and in the universal gadget, allows us to formulate a generalization that is even applicable in quantum computer circuits.

\section*{Acknowledgements}

BFR and VGV are in debt with Prof. Wallon A. T. Nogueira, Profa. Giovana Trevisan Nogueira and Prof. Rodrigo A. Dias for helpful discussions and recommendations for completing this work. CD wishes to thank the hospitality of both the Universidade Federal de Pernambuco and the Universidade Federal de Juiz de Fora, where part of this work was realized. LLB would like to express his gratitude to the Universidade Federal de Juiz de Fora and Programa de Bolsas de Pós-Graduação - Física, for his master's scholarship.

This work has also been supported by Programa Institucional de Bolsas de Iniciação Científica – VI VIC/Universidade Federal de Juiz de Fora – 2022/2023, project number 51268.

\appendix

\section{Basic set theory and the action of groups on arbitrary sets}
\label{appendixA}

In this appendix, we briefly introduce a selection of topics concerning set theory: equivalence relations, equivalence classes, the action of a group on a set and related concepts. Although the points we address here are not new, our intention is to make the manuscript as self-consistent as possible. For an in-depth description, we direct the reader to \cite{jauch_foundations_1973, nakahara_geometry_1990}. 

We start our overview by reviewing an alternative way of turning elements of a set into equivalent elements. Presumably, the first and foremost notion of equivalence is the one commonly connected with the concept of equality. Disguised by common sense, equality is a binary relation—a relation between two objects—which is reflexive (an object is equal to itself), symmetric (if an object is equal to another, then the latter is also equal to the former) and transitive (if an object is equal to a second, and this second to a third, then the original and the final objects also are so). What is crucial here is that equality, as we know it, is only a special case of a larger class of binary relations intended to capture and classify distinctive characteristics of a collection of objects. The next definitions rigorously address these points.

\begin{definition}
(Equivalence Relations).
Let $\mathds{X}$ be a non-empty set and  $\sim \subset \mathds{X} \times \mathds{X}$ a relation. We say that $\sim$ is an \emph{equivalence relation} whenever the three properties below hold true,

\textit{i.} $\sim$ is reflexive: $x \sim x$, $\forall x \in \mathds{X}$.

\textit{ii.} $\sim$ is symmetric: $x \sim y \Rightarrow y \sim x$, $\forall x,y \in \mathds{X}$

\textit{iii.} $\sim$ is transitive: if $x \sim y$ and $y \sim z$, then $x \sim z$, $\forall x,y,z \in \mathds{X}$.  
\end{definition}

\textbf{Remark.} Any relation over a given set $\mathds{X}$ is a subset $R$ of $\mathds{X} \times \mathds{X}$. In this sense, $R$ is a collection of ordered pairs $(x,y)$ where $x,y \in \mathds{X}$. Mainly because we want to express the relational aspect of $R$ and also retain the parallel with the notion of equality, it is usual to write $x R y$ instead of $(x,y) \in R$—the reader will certainly appreciate this change in notation.

\begin{definition}
(Equivalence Classes). Let $\sim$ be an equivalence relation over $\mathds{X}$. Given an element $y$  we define the following subsets of $\mathds{X}$,
\begin{equation}
    [y] := \{x\in \mathds{X} \, \vert \, x \sim y \}.
\end{equation}
Because $y \sim y$, $y \in [y]$, for all elements of $\mathds{X}$. Thus, these subsets are well-defined and we name them \emph{equivalence classes}. In particular, for a given $y \in \mathds{X}$, the subset $[y]$ is called the \emph{equivalence class of $y$}.
\end{definition} 

One of the main consequences of establishing an equivalence relation on a set resides in the theorem below. It guarantees that equivalence classes provide a partition of the set they are defined over---that is, a cover of the entire set formed by disjoint subsets.

\begin{theorem}\label{theorem.partition.equivalence}
(Partition via Equivalence Classes). If $\sim$ is an equivalence relation on $\mathds{X}$, then
\begin{eqnarray}
1) \, x \sim y \Rightarrow [x] = [y], 
\end{eqnarray}
\begin{eqnarray}
2) \, x \nsim y \Rightarrow [x] \cap [y] = \emptyset, 
\end{eqnarray}
\begin{eqnarray}\label{partition} 
3)\, \mathds{X} = \bigcup_{x \in \mathds{X}} [x].
\end{eqnarray}    
\end{theorem}

\textbf{Remark.} A useful notation to the disjoint union in eq. \eqref{partition} is $\mathds{X}/\sim$, as it denotes that the set has been partitioned across the many equivalence classes defined by the equivalence relation. For both an intuitive interpretation and formal demonstration, we direct the reader to \cite{vasconcelos_junior_grandezas_unidimensionais_2018}.  

Now we turn our attention to group actions on sets. Following the natural steps of introducing sets and relations, we could start with functions (or mappings), which are a special type of relation. However, we would like to do so bearing in mind that the elements of the set should transform guided by a group structure, retaining not only the symmetry properties but also the special transformations the latter usually conveys. Hence, we define the    
\begin{definition}
    (Action of a Group). Let $\mathcal{G}$ be a group and $\mathds{X}$ be a non-empty set. The \emph{action} of $\mathcal{G}$ on $\mathds{X}$ is a map $\varphi: \mathcal{G} \times \mathds{X} \rightarrow \mathds{X}$, satisfying the following conditions,

\textit{i.} $\varphi(e,x) = x$,  $\forall \,\, x \in \mathds{X}$, where $e$ stands for the group identity element.

\textit{ii.} For each $g \in \mathcal{G}$, $\varphi(g, \cdot): \mathds{X} \rightarrow \mathds{X}$ is a bijection. 

\textit{iii.} $\varphi(g_1, \varphi(g_2,x))=\varphi(g_1g_2,x)$, $\forall \,\, x \in \mathds{x}$ and $\forall \,\, g_1, g_2 \in \mathcal{G}$. 
\end{definition}

Now, in $\mathds{X}\times \mathds{X}$ we introduce the following relation,
\begin{equation}
    x \sim y \Leftrightarrow y = \varphi(g,x), \mbox{ for some } g \in \mathcal{G}.
\end{equation}
We affirm that $\sim$ is an equivalence relation. In fact, $\sim$ is reflexive because \textit{i}: $\varphi(e,x) = x$. The symmetry comes from 
$$y = \varphi(g,x) \Rightarrow x = \varphi(g^{-1},y). $$
Finally, the transitivity can also be promptly deduced. It suffices to note that if $y = \varphi(g_1,x)$ and $z = \varphi(g_2,y)$, then $z = \varphi(g_3,x)$, where $g_3 = g_2 g_1$.

With the equivalence relation defined by the action, we define the equivalence classes by 
\begin{equation}
    [x] := \{ y \in \mathds{X} \, \vert \,  y = \varphi(g,x) \mbox{ for some } g \in \mathcal{G} \}.
\end{equation}
Due to its special role, an equivalence class of such type is called an \textit{orbit} through $x$. 

Crucially, according to the theorem \ref{theorem.partition.equivalence} above, $\mathds{X}$ is partitioned by its disjoint orbits. All in all, that is to say, starting from the action of a group over a set, we can partition that set across the orbits of each of its elements.

Next, we will present some examples of interest not only to our previous discussion \textit{per se}, but also to quantum mechanics in general. 

\textit{Example 1. }For our first example, we consider $\mathds{X} = \mathds{C}^2$ and $\mathcal{G} = \mathcal{U}(1) = \{ e^{i \alpha} \, \ver \, \alpha \in \mathds{R} \}$. Define the map
\begin{align*}
    \varphi: \mathcal{U}(1) \times \mathds{C}^2 & \rightarrow \mathds{C}^2   \\
    (e^{i\alpha}, \kp ) &\mapsto \kp' = e^{i \alpha}\kp.
\end{align*}

To check that $\varphi$ is indeed an action, we have

\textit{i.} $\mathcal{U}(1) \ni e =1 \Rightarrow \varphi(e, \kp) = \kp$.

\textit{ii.} Given $\alpha \in \mathds{R}$, $\varphi(e^{i \alpha}, \cdot): \mathds{C}^2 \rightarrow \mathds{C}^2$ is a bijection. In effect, 
$$e^{i \alpha} \ket{\psi_1} = e^{i \alpha} \ket{\psi_2} \Rightarrow \ket{\psi_1} = \ket{\psi_2}$$
which guarantees that the map is injective. In turn, we point out that any $\kp \in \mathds{C}^2$ is reached by $e^{-i\alpha}\kp$ under $\varphi(e^{i \alpha}, \cdot)$. Thus, the map is surjective as well. 

\textit{iii.} At last, 
$$ \varphi(e^{i \alpha_1},e^{i \alpha_2} \kp) = e^{i(\alpha_1+ \alpha_2)}\kp = \varphi(e^{i \alpha_1}e^{i \alpha_2}, \kp),$$
which concludes the proof. 

The importance of the case in point lies on the fact that the orbits consists of indistinguishable states for two-level quantum systems as well as fibers in the Hopf fibration, as previously discussed  in section \ref{Sec.RelationPSandSS}. 

\textit{Example 2. }Our second example is given by $\mathds{X}= \mathds{R}^n$ and $\mathcal{G} = \mathds{R}^* = \mathds{R}\smallsetminus \{0 \}$, where the group product is the usual multiplication of non-zero real numbers. Define
\begin{align*}
    \varphi:\mathds{R}^* \times \mathds{R}^n & \rightarrow \mathds{R}^n   \\
    (\lambda, \Vec{v} ) &\mapsto \lambda \Vec{v}.
\end{align*}
It is not difficult to conclude that $\varphi$ is an action. In fact, we have,

\textit{i.} $\mathds{R}^* \ni e = 1 \Rightarrow \varphi(1,\Vec{v}) = \Vec{v}$.

\textit{ii.} Given $\lambda \in \mathds{R}^*$, $\varphi(\lambda, \cdot)$ is a bijection for 
$$\lambda \Vec{v}_1=\lambda \Vec{v}_2 \Rightarrow \Vec{v}_1=\Vec{v}_2,$$ 
which shows the injection. Moreover, any $\Vec{v} \in \mathds{R}^n$ can be obtained by applying $\varphi(\lambda, \cdot)$ to $\lambda^{-1}\Vec{v}$. Hence, the map is also surjective.   

\textit{iii.} Finally, $\varphi(\lambda_1, \lambda_2 \Vec{v}) = \varphi(\lambda_1 \lambda_2 \Vec{v})$.

The classes, or orbits, here are straight lines crossing the origin, although $\Vec{0} \neq [\Vec{v}]$, for every non-zero vector $\Vec{v}$ in $\mathds{R}^n$. Their union forms what is called the projective space, named $\mathds{R}\mathds{P}^{n-1}$. The generalization for $\mathds{C}^n$ and, accordingly, to $\mathds{C}\mathds{P}^{n-1}$ is straightforward. The value of the latter to quantum mechanics stems from its intrinsic connection to entanglement \cite{bengtsson_cp_2002}.

\end{document}